\documentclass{sf2a-conf2018}
\usepackage{graphicx}
\usepackage{hyperref}
\usepackage[]{natbib}  
\usepackage{epstopdf}
\usepackage{mathabx}

\def\BibTeX{{\rm B\kern-.05em{\sc i\kern-.025em b}\kern-.08em
    T\kern-.1667em\lower.7ex\hbox{E}\kern-.125emX}}
\bibpunct{(}{)}{;}{a}{}{,}  

\begin{document}

\TitreGlobal{SF2A 2017}


\title{Tidal dissipation in deep oceanic shells: \\ from telluric planets to icy satellites}

\runningtitle{Tidal dissipation in deep oceanic shells}

\author{P. Auclair-Desrotour}\address{Laboratoire d’Astrophysique de Bordeaux, Univ. Bordeaux, CNRS, B18N, allée Geoffroy Saint-Hilaire, 33615Pessac, France}

\author{S. Mathis$^{2,}$}\address{Laboratoire AIM Paris-Saclay, CEA/DRF - CNRS - Université Paris Diderot, IRFU/DAp Centre de Saclay, 91191 Gif-sur-Yvette, France}
\address{LESIA, Observatoire de Paris, PSL Research University, CNRS, Sorbonne Universités, UPMC Univ. Paris 06, Univ. Paris Diderot, Sorbonne Paris Cité, 5 place Jules Janssen, 92195 Meudon, France}

\author{J. Laskar}\address{IMCCE, Observatoire de Paris, CNRS UMR 8028, PSL, 77 Avenue Denfert-Rochereau, 75014 Paris, France}



\author{J. Leconte$^1$}

\setcounter{page}{237}


\maketitle


\begin{abstract}
Oceanic tides are a major source of tidal dissipation. They are a key actor for the orbital and rotational evolution of planetary systems, and contribute to the heating of icy satellites hosting a subsurface ocean. Oceanic tides are characterized by a highly frequency-resonant behavior, which is mainly due to the propagation of surface gravity waves in the case of thin oceans, and internal waves when they are deeper. In this work, we derive self-consistent ab initio expressions of the oceanic tidal torque as a function of the key physical parameters of the system (the ocean depth, the Brunt-V\"ais\"al\"a stratification frequency, the rotation rate, the tidal frequency, the Rayleigh friction). These solutions include the coupled mechanisms of internal and surface gravito-inertial waves, which allows us to study the case of planets hosting deep oceans and offer interesting prospects for the coupling between subsurface oceans and ice shells in the case of icy satellites. 
\end{abstract}



\section{Introduction}
 
Oceanic tides are responsible for $\sim$95 \% of the total energy generated by the Lunar semidiurnal tide \citep[e.g.][]{Lambeck1977} in spite of the negligible thickness of the Earth ocean compared to the Earth radius \citep[typically, $H \approx 6 \times 10^{-4}~R_\Earth$; e.g.][]{ES2010}. This shows evidence of the necessity to take into account the potential existence of oceanic layers in the characterization of recently discovered terrestrial planets such as those hosted by the TRAPPIST-1 ultra-cool dwarf star \citep[][]{Gillon2017,Grimm2018}. More specifically, it is crucial to characterize the impact of oceanic tides on the planetary rotation and orbital evolution to better constrain the history and evolution of these planets. As oceans are generally treated as thin layers \citep[typically through the so-called shallow water approximation; see e.g.][]{Webb1980}, the effects induced by their internal structure are rarely considered. Yet, although these effects are negligible in the case of the Earth, they could play a more important role for planets hosting potentially deep oceans, such as TRAPPIST-1 terrestrial planets, which are likely to have conserved an important part of their initial water reservoir \citep[][]{Bolmont2017}.

Following early studies \citep[e.g.][]{Tyler2011}, we developed an ab initio modeling of oceanic tides based upon the classical linear approach and taking into account both the oceanic stratification and friction with the oceanic floor in a self-consistent way. We computed from this model analytic solutions expressing the oceanic tidal torque and Love numbers as explicit functions of the tidal frequency and key parameters (ocean depth, Rayleigh drag coefficient, Brunt-V\"ais\"al\"a frequency). This work is detailed in \cite{ADMLL2018} and we succinctly summarize here its main results by showing that they can be adapted to the study of icy satellites hosting subsurface oceans.

\section{Frequency dependence of the oceanic tidal torque}

\begin{figure}[ht!]
 \centering
 \includegraphics[height=0.26\textheight,trim = 1.5cm 2.0cm 6.5cm 2.3cm,clip]{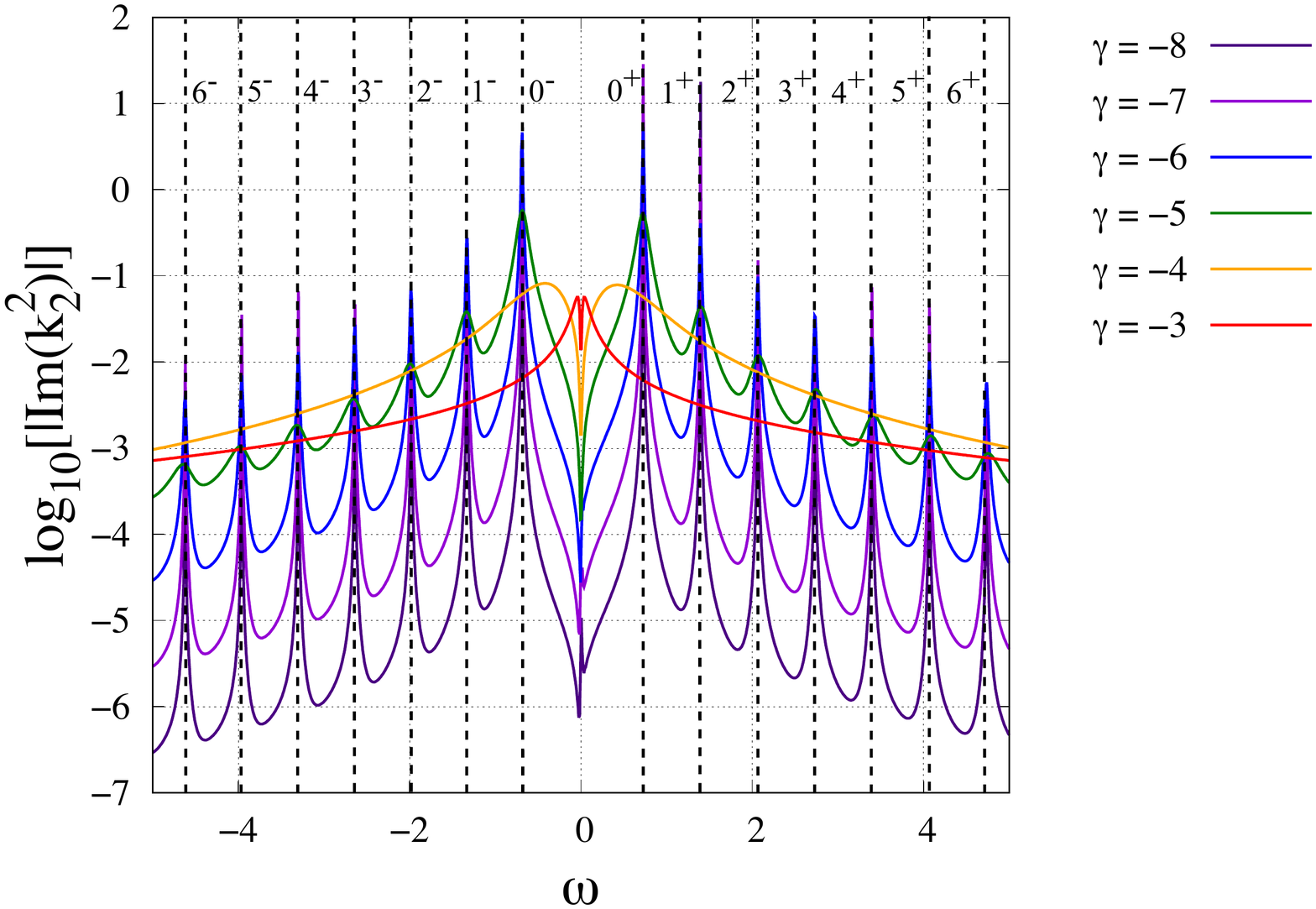}  \hspace{0.5cm}
  \includegraphics[height=0.26\textheight,trim = 1.5cm 2.0cm 1.5cm 2.3cm,clip]{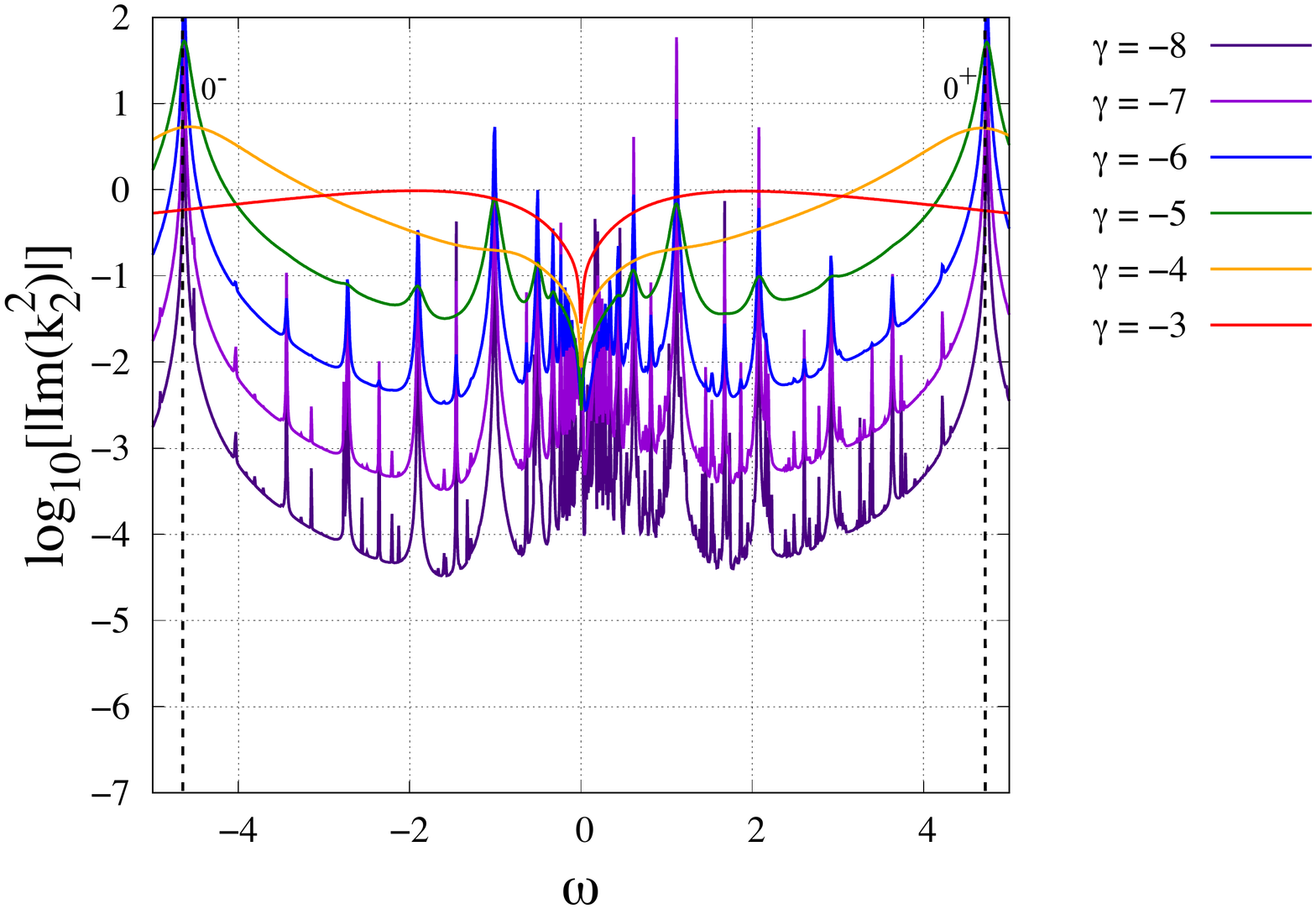}       
  \caption{Imaginary part of the quadrupolar Love number associated with the oceanic semidiurnal tide as a function of the normalized tidal frequency $\omega = \left( \Omega - n_{\rm orb} \right) / \Omega_\Earth$ (the notation $\Omega_\Earth$ designating the today rotation rate of the Earth) for various orders of magnitude of the Rayleigh drag coefficient $\gamma = {\rm log} \left( \sigma_{\rm R} \right) $. {\bf Left:} Earth-like ocean planet with a uniform ocean 4~km deep. {\bf Right:} Idealized TRAPPIST-1~f planet with a uniform ocean 1000~km deep. In each case, the orbital frequency is assumed to be constant and the rotation rate of the planet is related to the semidiurnal tidal frequency $\sigma$ through the formula $\sigma = 2 \left( \Omega - n_{\rm orb} \right)$. Resonances associated with surface inertia-gravity modes are designated by black dashed lines and numbers indicate the degree n of the corresponding Hough modes and the sign of their eigenfrequencies.}
  \label{auclair-desrotour1:fig1}
\end{figure}

Consider a terrestrial planet of radius $R_{\rm p}$ hosting an ocean of uniform depth $H$, density $\rho_{\rm s}$ at the surface, rotating at the angular velocity $\Omega$, and undergoing gravitational tides generated by a given perturber, star or satellite. In the case where the perturber orbits the planet circularly in its equatorial plane (the Keplerian orbital frequency is denoted $n_{\rm orb}$), the second order Love number describing the quadrupolar distortion of the layer can be written as 

\begin{equation}
k_2^2 = \frac{G M_{\rm oc}}{5 R_{\rm p}} \sum_n C_{2,n,2}^{2,\nu}  \mathcal{Q}_n^{2,\sigma},
\end{equation}

\noindent where we have introduced the gravitational constant $G$, the ocean mass in the shallow water approximation $M_{\rm oc} = 4 \pi R_{\rm p}^2 H \rho_{\rm s} $, the semidiurnal tidal frequency $\sigma = 2 \left( \Omega - n_{\rm orb} \right)$, the latitudinal wavenumbers of Hough modes $n$, and the associated Coriolis coupling coefficients $C_{2,n,2}^{2,\nu}$ and components of the quadrupole moment $\mathcal{Q}_n^{2,\sigma}$. For a uniform oceanic stratification with respect to convection (i.e. a uniform Brunt-V\"ais\"al\"a frequency), the $\mathcal{Q}_n^{2,\sigma}$ can be expressed as explicit functions of the tidal frequency and physical system parameters. This analytic solution is plotted in Fig.~\ref{auclair-desrotour1:fig1} for two asymptotic cases: an Earth-like ocean planet with a 4~km deep incompressible ocean (left panel), and an idealized TRAPPIST-1~f planet with a 1000~km deep stably-stratified compressible ocean (right panel). In the first case, the spectrum of the tidal torque is shaped by resonances resulting from the propagation of surface gravito-inertial waves associated with Hough modes. When stable-stratification is taken into account, internal gravity waves can propagate, leading to the resonances observed in the second case. 

\section{Application to icy satellites hosting subsurface oceans}

\begin{figure}[ht!]
 \centering
 \includegraphics[width=0.48\textwidth,clip]{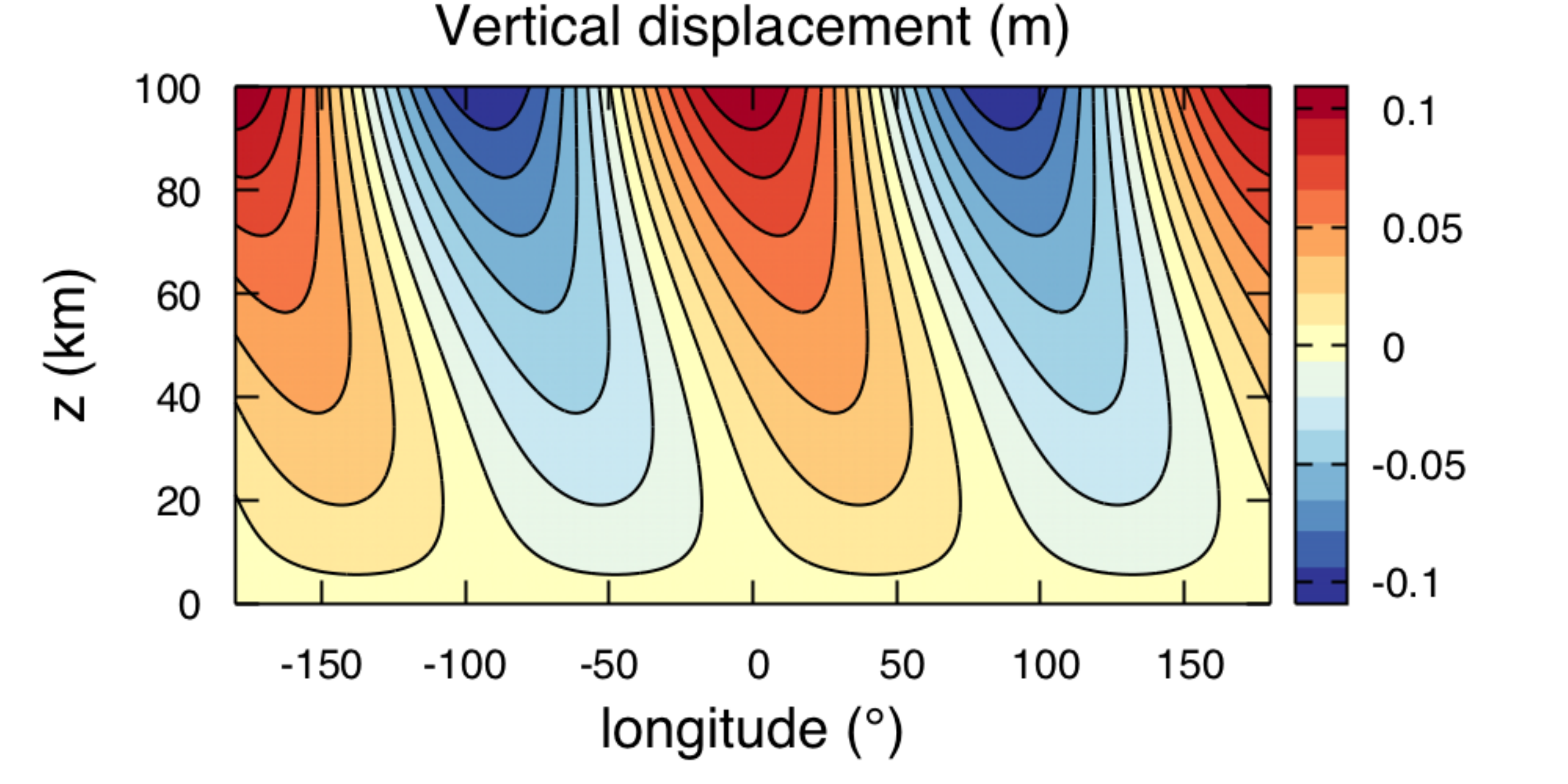}  
  \includegraphics[width=0.48\textwidth,clip]{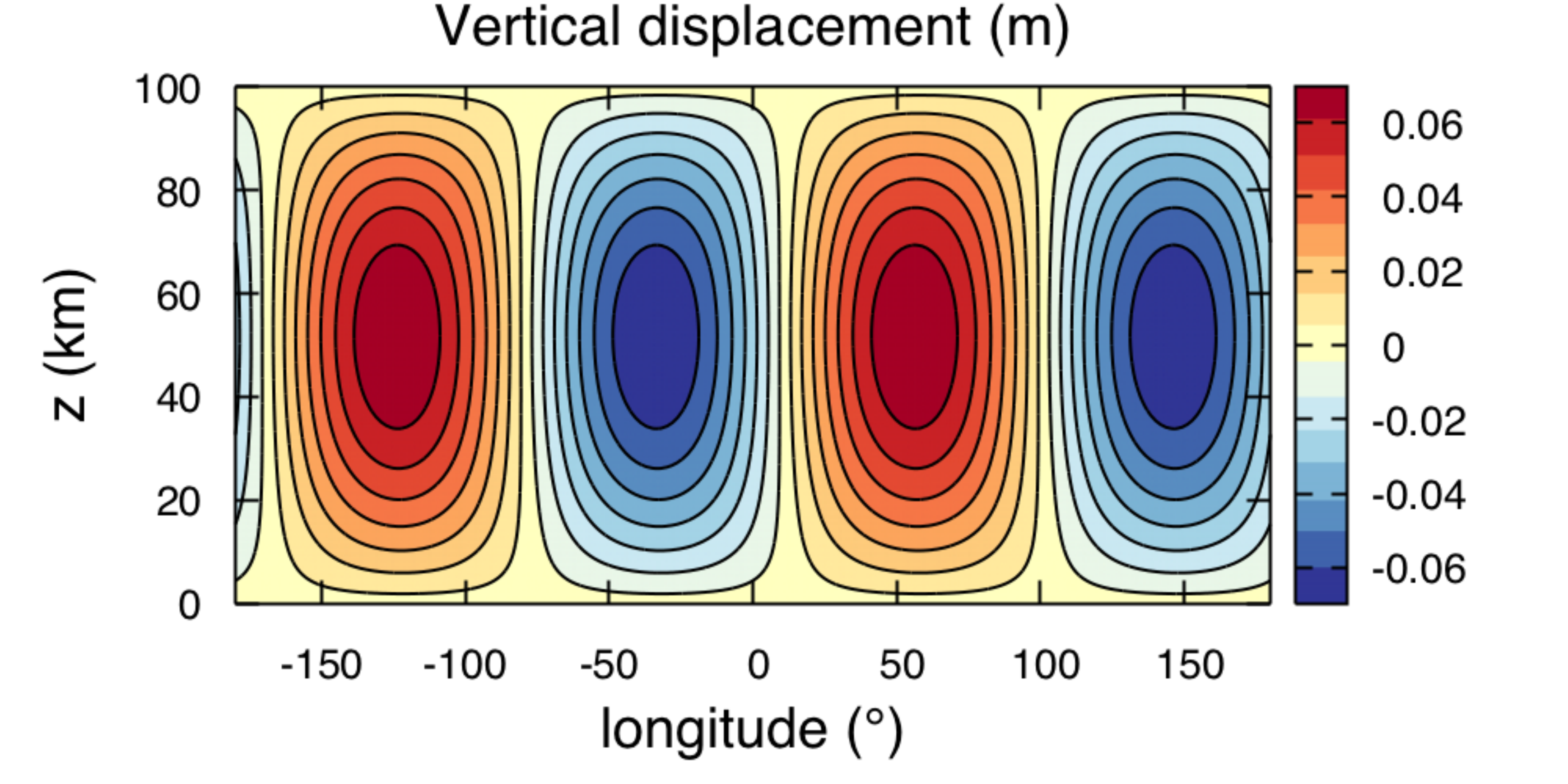}       
  \caption{Vertical displacement created by the semidiurnal tide in the equatorial plane of a 100~km deep global ocean as a function of longitude ($^\degree$) and altitude (km). {\bf Left:} Planetary ocean (standard free-surface boundary condition). {\bf Right:} Subsurface ocean (rigid lid). In both panels, the response is computed from the analytic solution derived in the case of a uniformly stably-stratified fluid layer (i.e. the Brunt-V\"ais\"al\"a frequency is taken constant) and for the unitary quadrupolar tidal potential $U_2^2 = 1 \ {\rm m^2.s^{-2}}$.}
  \label{auclair-desrotour1:fig2}
\end{figure}

In the Solar system, a variety of observations suggests the existence of subsurface oceans in an important fraction of satellites orbiting giants planets, such as Enceladus, Europa, Ganymede and Callisto \citep[e.g.][]{Kivelson2000,Zimmer2000}. As advanced models developed to quantify the resulting internal tidal heating of these bodies generally assume an incompressible oceanic layer \citep[e.g.][]{Matsuyama2018}, the solution derived above for ocean planets can be used as a first approximation to investigate the role played by the stratification of subsurfaces ocean in the tidal response. This point is illustrated by Fig.~\ref{auclair-desrotour1:fig2} where the vertical displacement created by the semidiurnal tide in the equatorial plane of a 100~km deep ocean is plotted as a function of longitude and altitude in two cases. The standard free-surface condition used for planetary oceans (left panel) is replaced by a rigid lid in the case of icy satellites (right panel), which filters surface gravity waves and reduces the response to the contribution of internal gravity waves.

\section{Conclusions}

In order to better understand the impact of the ocean internal structure on the tidally generated oceanic energy dissipation, we calculated an analytic solution describing the tidal response of a uniformly stably-stratified fluid layer. We used this solution to explore the parameter space, characterize the frequency-behaviour of the oceanic tidal torque, and provide a diagnosis about the nature of waves generating resonances in the tidal response. We showed that the obtained solution may also be used to examine how stratification may affect the tidal heating of subsurface oceans hosted by icy satellites.

\begin{acknowledgements}
The authors acknowledge funding by the European Research Council through ERC grants SPIRE 647383 and WHIPLASH 679030, the Programme National de Planétologie (INSU/CNRS) and the CNRS PLATO grants at CEA/IRFU/DAp. 
\end{acknowledgements}

\bibliographystyle{aa}  
\bibliography{auclair-desrotour1} 

\end{document}